\documentclass[twocolumn,showpacs,preprintnumbers,amsmath,amssymb]{revtex4}
\usepackage{graphicx}
\usepackage{dcolumn}
\usepackage{bm}
\usepackage{soul,color}
\begin{document}
\title{Reducing Financial Avalanches By Random Investments}

\author{Alessio Emanuele Biondo $^1$, Alessandro Pluchino $^2$, Andrea Rapisarda$^2$, Dirk Helbing $^{3}$}
\bigskip 

\affiliation{$^1$ Dipartimento di Economia e Impresa - Universit\'a di Catania, Corso Italia 55, 95129 Catania, Italy\\
$^2$ Dipartimento di Fisica e Astronomia, 
Universit\'a di Catania  and INFN sezione di Catania, 
Via S. Sofia 64, 95123 Catania, Italy\\
$^3$ ETH Zurich, Clausiustrasse 50, 8092 Zurich, Switzerland  }
\date{\today}
\bigskip
\begin{abstract}
\noindent Building on similarities between earthquakes and extreme financial events, we use a self-organized criticality-generating model to study herding and avalanche dynamics in financial markets. We consider a community of interacting investors, distributed on a small-world network, who bet on the bullish (increasing) or bearish (decreasing) behavior of the market which has been specified according to the S\&P500 historical time series. Remarkably, we find that the size of herding-related avalanches in the community can be strongly reduced by the presence of a relatively small percentage of traders, randomly distributed inside the network, who adopt a random investment strategy. Our findings  suggest a promising strategy to limit the size of financial bubbles and crashes. We also obtain that the resulting wealth distribution of all traders corresponds to the well-known Pareto power law, while the one of random traders  is exponential. In other words, for technical traders, the risk of losses is much greater than the probability of gains compared to those of random traders. \end{abstract}

\pacs{89.65.Gh,89.65.Gh, 05.65.+b 
}
\maketitle
\vspace{0.25cm}

Financial markets often experience extremes, called ``bubbles'' and ``crashes''. The underlying dynamics is related to avalanches, the size of 
which is distributed according to power laws \cite{Mandelbrot,StanleyMantegna,PeinkeNature,Farmer1,Bouchaud,Sornette,Helbing-Nature,Buchanan}. Power laws imply that crashes may reach any size---a circumstance that may threaten the functionality of the entire financial system. Many scientists see ``herding behavior'' as the origin of such dangerous avalanches \cite{AkerlofAnimalSpirits,BehavioralEconomicsReferences,Hommes,Shapira,Preis1,Preis2,HelbingHolystPRL,ParisiSornetteHelbingPRE}. In our paper, we explore whether there is a mechanism that could stop or reduce them. 
\par
To generate a  power-law dynamics similar  to the volatility clustering in financial markets, we use an agent-based model that produces the phenomenon of self-organized criticality (SOC) \cite{Bak}. Specifically, we adapt the Olami-Feder-Chrstensen (OFC) model \cite{Olami,Caruso} that has been proposed to describe the dynamics of earthquakes \cite{Sornette,Mantegna1}. In this context, we assume information cascades between agents \cite{Bikhchandani-Hirshleifer-Welch} as the underlying mechanism of financial avalanches. We assume that agents interact within a small-world (SW) network of financial trading \cite{Caruso3} and that there is social influence among them \cite{GrundWaloszekHelbing,Economics2.0}. Such kinds of models are recently becoming popular in economics \cite{Tedeschi} and herding effects are also beginning to be observed in lab experiments \cite{Helbing-Yu}. 
In this connection, it is interesting to consider that humans tend to orient themselves at decisions and behaviors of others, particularly in situations where it is not clear what is the right thing to do \cite{HelbingVicsek}. Such conditions are typical for financial markets, in particular during volatile periods. In fact, in situations of high uncertainty, personal information exchange may reach market-wide impacts, as the examples of bank runs \cite{diamond-dybvig} and speculative attacks on national currencies  \cite{eichengreen-rose-wyplosz} show. 
\par
Our study explores how huge herding avalanches in financial trading might be reduced by introducing a certain percentage of traders who  adopt a random  investment strategy. Actually, several analogies between socio-economic and physical or biological systems have recently been discussed \cite{Stanley, Gammaitoni, CarusoHuelga, Thurner}, where noise and randomness can have beneficial effects,  improving the performance of the system \cite{Pluchino1, Pluchino2, Pluchino3, Sornette1, Sornette2, Schwartz,Helbing-Yu2,Farmer,Taleb}. More specifically, in a recent series of papers, it has been explored whether the adoption of random strategies may be advantageous in financial trading from an individual point of view. Scientific evidence suggests that, in the long term, random trading strategies are less risky for a single trader, but provide, on average, gains comparable to those obtained by technical strategies \cite{Biondo1, Biondo2}. 
Therefore, one might expect that a certain percentage of investors would consider the possibility of adopting a random trading strategy. Assuming  this and  using real data sets of the S\&P500 index, we here extend  our previous analysis to a sample community of interacting investors. We  investigate whether the presence of randomly distributed agents performing random investments  influences the formation of herding-related avalanches and how the wealth  of  the traders is distributed. We show that the presence  of random traders is able to reduce financial avalanches, which we call - in analogy with earthquakes - 'financial quakes'. Furthermore, we find that the wealth distribution, even if normally distributed in the beginning, spontaneously evolves towards the well-known Pareto power law.
Finally, we  address possible policy implications.
\begin{figure}
\begin{center}
\includegraphics[width=3.0in,angle=0]{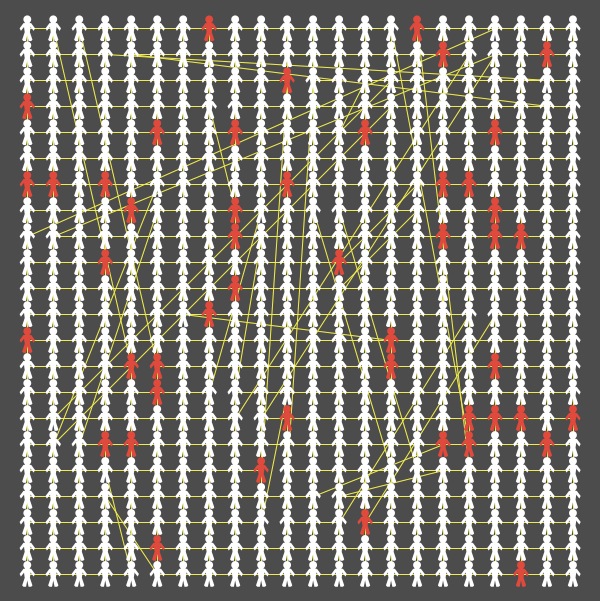}
\caption{\label{fig1} 
Small-world $2D$ network of $N$ traders, as used in our numerical simulations (here $N=500$ as an example, but in the simulations we always considered $N=1600$ agents). Short-range and long-range links are simultaneously present. White agents are RSI traders (active agents  assumed to follow the standard Relative Strength Index (RSI) trading strategy); colored agents ($10\%$ of the total) are random traders, here uniformly distributed at random among the population. See text for further details.}
\end{center}
\end{figure}

\section{The Model} 

The Financial Quakes Model (FQM) that we introduce in this paper is defined on a small-world (SW) network \cite{Caruso3} of $N$ agents  $A_i$ ("traders").  The total number of traders considered is always $N=1600$. Note that the purpose of this model is to generate volatility clustering and power-law distributed avalanche sizes in a simple way, while we are not interested in formulating a realistic micro-model of financial markets or to reproduce the exact power-law exponents. The purpose of our study is to explore possible ways to destroy dangerous herding effects  by simple and effective means.
\par
In our FQM model, each agent carries a given quantity of information about the financial market considered. The SW network is obtained from a square 2-dimensional $40 \times 40$ lattice with open boundary conditions, by randomly rewiring the nearest neighbors links with a probability of $p=0.02$ (see Fig.1). The resulting network topology allows the information to spread over the lattice through long-range links, but also preserves the clustering properties of the network and its average degree ($<k>=4$). 
\par       
The information spreading is simulated by associating to each trader a real variable $I_i(t)$ $(i=1,2,...,N)$, representing the information possessed at time $t$, which initially (at $t=0$) is set to a random value in the interval $(0,I_{th})$. $I_{th}=1.0$ is a threshold value that is assumed to be the same for all agents. At each discrete time step $t>0$, due to public external information sources, all these variables are simultaneously increased by a quantity $\delta I_i$, which is different for each agent and randomly extracted within the interval $[0,(I_{th}-I_{max}(t))]$, where $I_{max}(t)= \max \{I_i(t)\}$ is the maximum value of the agents' information at time $t$. 
\par
\begin{figure}
\begin{center}
\includegraphics[width=3.4in,angle=0]{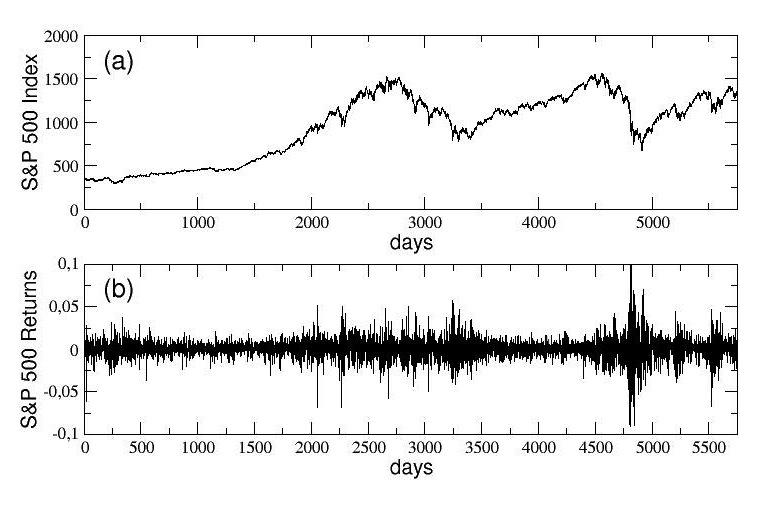}
\caption{\label{fig2} 
(a) S$\&$P 500 index data used in this paper, with $5750$ daily index entries $E_j$ (from September 11, 1989, to June 29, 2012). (b) Corresponding (relative) returns time series, where returns are defined as the ratio $(E_{j+1}-E_j)/E_j$.}
\end{center}
\end{figure}
If, at a given time step $t^*$, the information $I_k(t^*)$ of one or more agents $\{A_k\}_{k=1,...,K}$ exceeds the threshold value $I_{th}$, these agents become "active" and take the decision of investing a given quantity of money by betting on the bullish (increasing) or bearish (decreasing) behavior of the market compared to the day before. As mentioned before, we consider here as a typical example the S$\&$P $500$ index. The time period ranges  from September 11, 1989, to June 29, 2012, over  a total of $T=5750$ daily index values $E_j$ (see Fig. 2).  Notice that the use of this particular series has no special reasons, it just serves to ensure  only a realistic  market dynamics as input. Other indexes have also been tested with similar results. 

In order to make their prediction $P_j$ (positive or negative) about the sign of the index difference $(E_j-E_{j-1})$ at time $t^*$, active agents are assumed to follow the standard Relative Strength Index (RSI) trading strategy, based on the ratio between the sum of positive returns and the sum of negative returns experienced during the last $\tau_{RSI}$ days (here we choose  $\tau_{RSI}=14$; see Ref. \cite{Biondo2} for further details of the RSI algorithm). As for the time series considered, this strategy has nothing special and has been chosen just because it is a commonly used technical strategy in the trading community. 

For financial traders, it is often beneficial to be followed by others, as this increases the likelihood that their investments will be profitable or because they are friends/colleagues and it would be considered appropriate  to share part of their own information. Therefore, we assume that the agents, once activated, will transfer some information to the neighbors according the following herding mechanism:
\begin{equation}   
\label{av_dyn}       
I_k > I_{th}  \Rightarrow \left\{ 
	\begin{array}{l}
       I_k \rightarrow 0, \\
       I_{nn} \rightarrow I_{nn} + \frac{\alpha}{N_{nn}} I_k .       \end{array} 
	\right.
\end{equation}
Here ``nn'' denotes the set of nearest-neighbors of the active agent $A_k$. $N_{nn}$ is the number of direct neighbors, and the parameter $\alpha$ controls the level of dissipation of the information during the dynamics ($\alpha=1$ corresponds to the conservative case). In analogy with the OFC model for earthquakes \cite{Caruso,Caruso3}, we set here $\alpha=0.84$ (non-conservative case), i.e. we consider some information loss during the herding process. 
This value of $\alpha$ has been  chosen here  to drive the system in a critical state and to obtain large avalanches, since our goal is to study how these avalanches can be reduced by the introduction of random traders.
\begin{figure}
\begin{center}
\includegraphics[width=3.55in,angle=0]{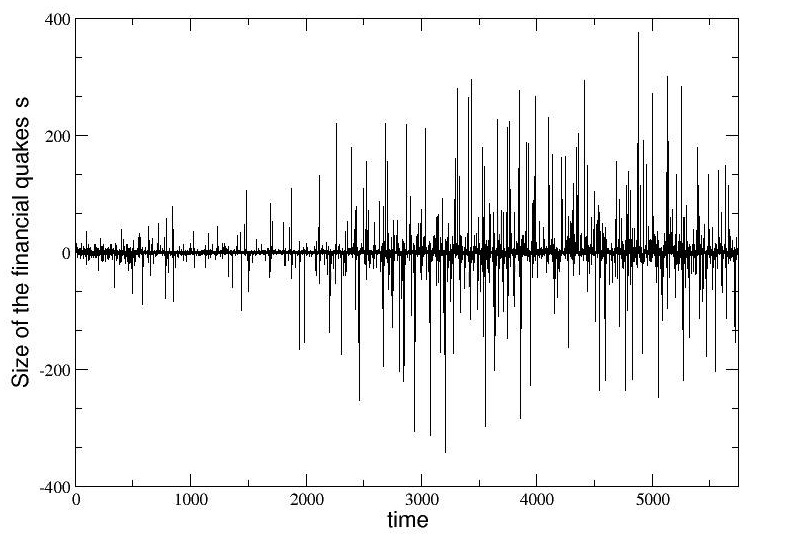}
\caption{\label{fig3} 
Size of avalanches (``financial quakes'') occurring in our artificial financial market. Both positive cascades (``bubbles'') and negative avalanches (``crashes'') are found. See text for further details.}
\end{center}
\end{figure}

Of course, the herding rule (\ref{av_dyn}) can activate other agents, thereby producing a chain reaction. The resulting information avalanche may be called a ``financial quake'': all the agents that are above the threshold become active and invest simultaneously according to Eq. (\ref{av_dyn}), such that the agent $A_k$  bets with the same prediction $P_j$ as the agent from which they have received the information. The financial quake is over, when there are no more active agents in the system (i.e. when $I_i < I_{th}$ $\forall i$). Then, the prediction $P_j$ is finally compared with the sign of $(E_j - E_{j-1})$: if they are in agreement, all the agents who have contributed to the avalanche win, otherwise they lose. In any case, the process of information cascades build up again due to the random public "information pressure" acting  on the system. The number of investments (i.e. the number of active agents) during a single financial quake define the avalanche size $s$. 

In the next section we present several simulation results obtained by running this model many times, starting each time from a new random initial distribution of the information $I_i(0)$ shared among the agents. 
We assume that the avalanche process within our sample trading community does not influence the whole market, i.e. does not have any effect on the behavior of the financial series considered, even if we imagine that the market does exert some influence on our community through "the information pressure" $\delta I_i$. On the other hand, this pressure, being random and different for each agent, is also  independent from the financial time series considered (here the $S\&P500$). 

In a way, this scenario could be considered analogous to the physical situation of a small closed thermodynamical system in contact with a very large energy {\it reservoir} (environment), typical for statistical mechanics in the canonical ensemble: even if the system can exchange energy with the environment, it is too small to have any influence on the reservoir itself.

\par\begin{figure}
\begin{center}
\includegraphics[width=3.55in,angle=0]{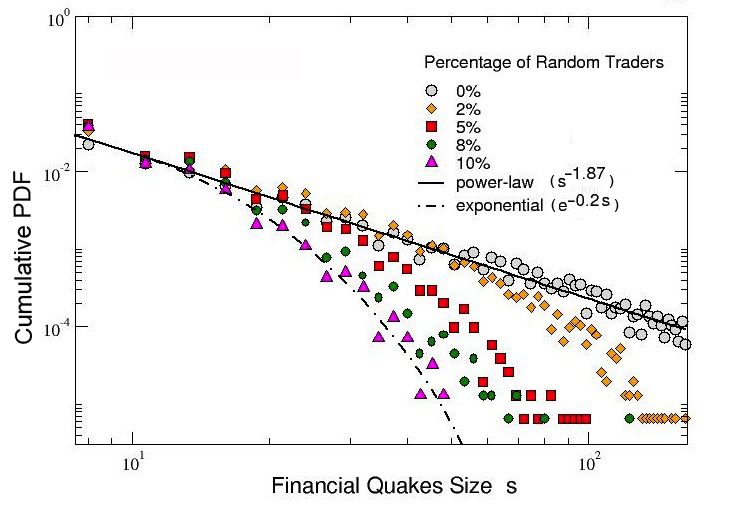}
\caption{\label{fig4} 
Cumulative distribution of the sizes (absolute values) of  herding avalanches occurring in the community of investors, with and without random traders. In the absence of random traders (open circles) he distribution obeys a power law with an exponent equal to -1.87 (a fit is also reported as a straight line).  Considering increasing amount of random traders, i.e. $2\%$ (diamonds), $5\%$ (squares), $8\%$ (full circles) and $10\%$ (triangles), the distribution tends to become exponential for sufficiently large percentage of random investors. An exponential fit with exponent equal to $-0.2$ is also found for the latter case. In these simulations the random traders are uniformly distributed (at random) over the network, as in Fig.1. For further details see the main text.}
\end{center}
\end{figure}

\section{Numerical results} 

In Fig.~3 we plot the time sequence $s(t)$ of the ``financial quake'' sizes during a single simulation run. A positive sign means that all the involved agents win and a negative sign that they loose. Each avalanche corresponds to an entry of the S\&P $500$ index series, since each initial investment (``bet'') on the market coincides with the occurrence of a financial quake (this means that the series in Fig.~2 and Fig.~3 have the same length $T$). In analogy with the SOC behavior characterizing the OFC model, we observe a sequence of quakes that increases in size over time. In other words, the financial system is progressively driven into a critical-like state, where herding-related avalanches of any size can occur: most of them will be quite small, but sometimes a very big financial quake appears, involving a herding cascade of bets, which can be either profitable (positive) or lossfull (negative).  
Notice that the daily data of the S\&P $500$ series only affect the sign of the avalanches in Fig.~3, while their sizes strictly depend on the internal dynamics of our small trading community considered. Therefore, as we verified with several simulations not reported here, reshuffling data or removing extreme events in the S\&P $500$ series would not produce any change in the sizes of the financial quakes.

The SOC-like nature of this dynamics is well shown in Fig.~4, where we report the probability distribution $P_N (s)$ of financial quakes size, measured by  its absolute value and cumulated over 10 simulations (open circles). The resulting  distribution can be very well fitted by a power-law $P_N (s) \propto s^{-1.87}$, a slope  consistent with the one  obtained for earthquakes in the OFC model on a SW topology (see \cite{Caruso3}).          
\par\begin{figure}
\begin{center}
\includegraphics[width=3.5in,angle=0]{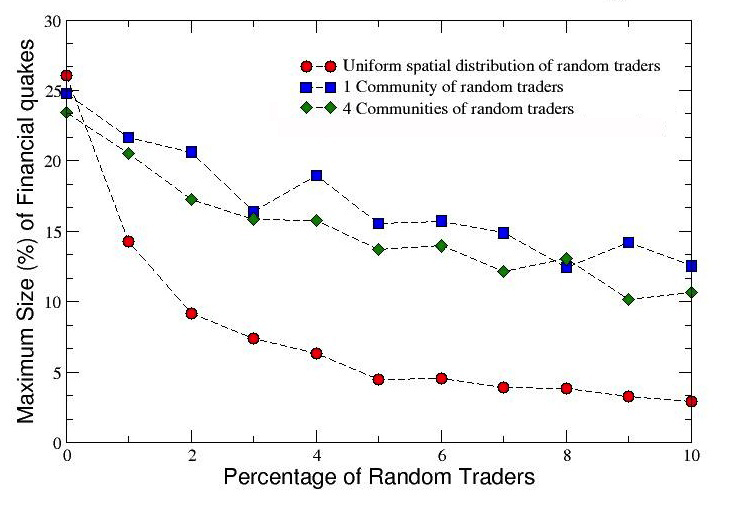}
\caption{\label{fig5} 
Behavior of the maximum size of avalanches as a function of an increasing number of random traders, for three different ways of spreading them over the network: randomly distributed  (circles), grouped in one community (squares) and grouped in four communities (diamonds). The results were averaged over 10 different realizations, each one with a different initial random position of traders and communities within the network.}
\end{center}
\end{figure}

Let us now discuss what happens if a certain number of agents in the network adopt a random trading strategy, i.e. if they invest in a completely random way instead of following the standard RSI strategy. 
We have already shown \cite{Biondo2} that an individual random trading strategy, if played along the whole S\&P 500 series (and also along other European financial indices), performs as well as various standard trading strategies (such as RSI, MACD or Momentum \cite{Biondo2}), but it is less risky than other strategies. Here, we study whether a widespread adoption of such a random investment strategy would also have a beneficial (collective) effect at the macro-level (where other important phenomena like herding, asymmetric information or rational bubbles may matter). Would random investment strategies reduce the level of volatility and induce a greater stability of financial markets?  
 
In the following, we test  this hypothesis by introducing a certain percentage $P_{RND}$ of random traders (colored agents in Fig.1), uniformly distributed at random among  the  $N=1600$ investors. We   assume that all agents are aware of the trading strategy (RSI or random) adopted by their respective neighbors. In this respect it is worthwhile to stress again that, in our model, traders behave according to a bounded rationality framework with no feedback mechanism on the market. Note that, in contrast to RSI traders,  random traders are not activated by their neighbors, since they invest at random. We also assume that they do not activate their neighbors, since a random trader has no specific information to transfer.  In other words, random traders only receive information from  external sources,  but do not exchange individual information with other agents apart from the fact that they bet at random. 
\begin{figure}
\begin{center}
\includegraphics[width=1.65in,angle=0]{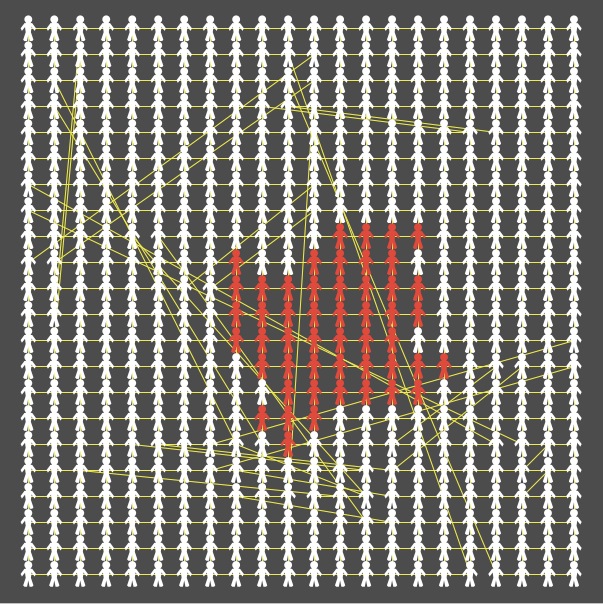}
\includegraphics[width=1.65in,angle=0]{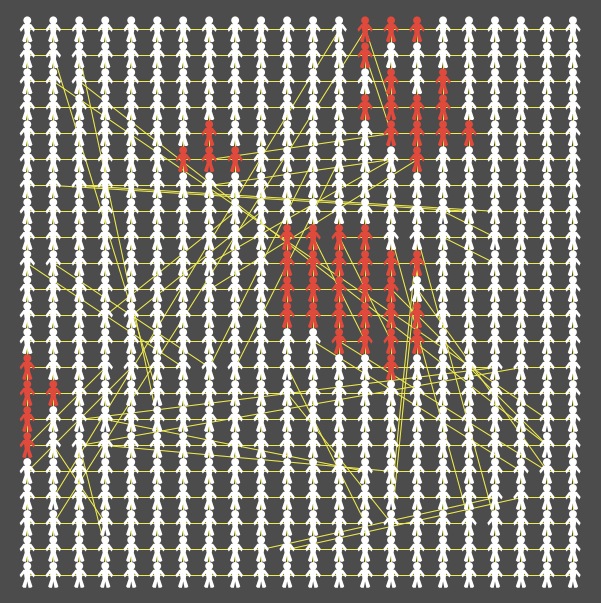}
\caption{\label{fig6} 
Two different examples of small-world $2D$ networks, with $N=500$ agents as in Fig. 1, but where random traders (colored agents, $10\%$ of the total) are grouped in one community (left panel) or four communities (right panel) , respectively.}
\end{center}
\end{figure}
In order to simulate random investors within our model, we simply set $\alpha = 0$ for them in Eq.~(\ref{av_dyn}). This means that random traders (when they  overcome their information threshold) can invest their capital exactly in the same way as other agents, but they do not take part in any herding-related activation avalanche. 

\par
Coming back again to Fig. 4, one can see the effect of an increasing percentage $P_{RND}$ of random traders on the size distribution of financial quakes. Besides the power-law curve already discussed, corresponding to $P_{RND}=0\%$, we report also the results obtained considering different percentages of random traders, when namely  $2\%$ (diamonds), $5\%$ (squares), $8\%$ (full circles) and $10\%$ (triangles). The data show that the original power-law distribution evolves towards an exponential one. An exponential fit with an exponent equal to -0.2 (dashed-dotted curve) is also reported for the maximum number of random traders considered by us, i.e. $10\%$. 

One can also investigate how the size of the avalanche changes with the increase of the amount of random traders considered, if they are uniformly distributed over the network. This is shown in Fig.5 (full circles), where one can see that  the maximum size of the avalanches observed drops by a factor of $5$ in the presence of only $5\%$ of random traders, reaching almost its final saturation level of $3\%$ when $P_{RND}=10\%$.
These results indicate that even a relatively small number of random investors distributed at random within the market is able to suppress dangerous herding-related avalanches. 
But what would happen if these random traders, instead of being uniformly distributed at random over the population, were grouped together in one or more communities? 

\begin{figure}
\begin{center}
\includegraphics[width=3.6in,angle=0]{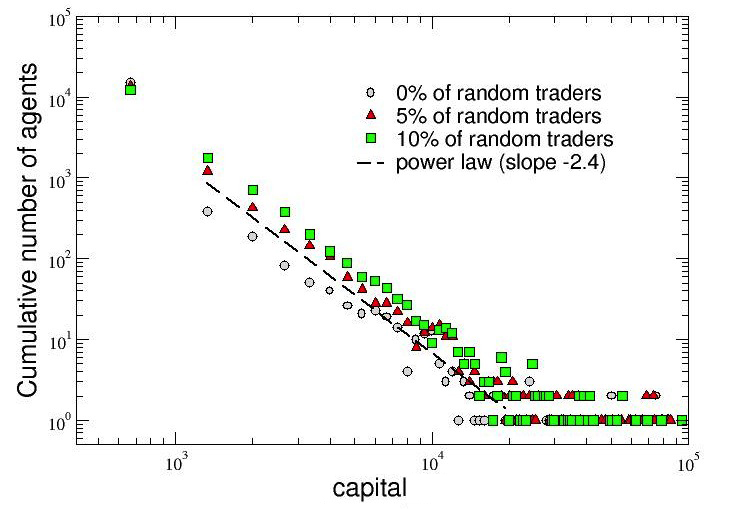}
\caption{\label{fig7} 
Capital/wealth distribution of all  agents at the end of the simulation, cumulated over 10 realizations, for a network with various percentages of uniformly distributed random traders. 
In any case we had a Pareto-like power law with an exponent of $-2.4$, independently of the amount of random traders considered.}
\end{center}
\end{figure}

In Fig.6 we show two examples of small-world networks with $10\%$ random traders (colored agents) grouped in one or four communities,  respectively (for clarity, we use the same sample network as in Fig.1, with $N=500$). If one repeats the previous simulations for our network of $N=1600$ agents with an increasing percentage of random traders, but now grouping these traders together in either one community or four communities, respectively, the result is that the original power law  distribution of avalanches is less affected by random investments for any percentage $P_{RND}$. This, in turn, implies a slower decrease of the maximum avalanche sizes as $P_{RND}$ increases, as shown again in Fig.5 (squares and diamonds, respectively). This means that the uniform random distribution of random investors over the whole network is quite crucial in order to significantly dampen  avalanche formation (notice that a percentage between $1\%$ and $2\%$ of uniformly distributed random traders is enough to reduce the financial quakes size as much as $10\%$ grouped random investors would do). In this respect, the effect of random traders is to increase the frequency of small financial quakes and, consequently, avoid the occurrence of large ones.     

It is also interesting to study the capital gain or loss, i.e. the change in wealth, of the agents involved in the trading process during the whole period considered (in the following we will use the terms "capital" and "wealth" synonymously). At the beginning of each simulation, we assign to each trader (RSI or random) an initial capital $C_i$ according to a normal distribution with an average of $<C>=1000$ credits and a standard deviation equal to $0.1 <C>$. Then we let them invest in the market according to the following rules:  

- if an agent wins thanks to a given bet (for example after being involved in a given, big or small, positive financial quake), in the next investment he will bet a quantity $\delta C_i$ of money equal to one half of his/her total capital $C_i$, i.e.  $\delta C_i=0.5C_i$;

- if an agent loses due to an unsuccessful investment (for example after a negative financial quake), the next time he will invest only ten percent of his/her total capital, i.e. $\delta C_i=0.1C_i$. We have checked, however, that our result are quite robust to adopting a number of different investment criteria.

After a financial quake, the capital of each agent involved in the herding-related avalanche will increase or decrease by the quantity $\delta C_i$. 
Of course random traders, who do not take part in avalanches, can invest. Their wealth   changes only when they overcome their information threshold due to the external information sources.
In Fig.7 we show the distribution of the total wealth cumulated by all  agents during the whole sequence of financial quakes, i.e. over the whole S\&P 500 series (cumulated over 10 different runs), for three different trading networks with increasing percentages of random investors (namely, $0\%$, $5\%$ and $10\%$). Interestingly, a Pareto power law \cite{pareto,bouchaud,garlaschelli} with an exponent equal to $-2.4$ (see the fit reported as dashed line) emerges spontaneously from the dynamics of the asymmetric investments, independently of the number of random traders. We have checked that this result is quite robust and does not substantially change if we modify the quantity $\delta C_i$  that agents choose  to invest  in case of a win or a loss.

It is also interesting to study  the wealth distribution of the random investors in  case of a network with $10\%$ random traders (results are cumulated over  $10$ realizations). This is reported in Fig.8, in comparison with the corresponding distribution already shown in the previous figure for the whole trading community. As one can see from the plot, random traders have a final wealth distribution very different from a power law, which can be  fitted very well with an exponential curve, represented by a dashed line, with an exponent equal to $-0.00134$ (the fact that the random traders component is not changing  the global power law distribution of Fig.7 is evidently due to its small size, i.e. $160$ agents as compared to  $1600$). In addition, we compare  average final wealth  of  all the traders,  767 credits, with the average wealth of random investors only, 923 credits. This   should be compared with the initial value of the capital which is  1000 credits. Therefore $14\%$ of RSI traders have more wealth in the end than in the beginning, whereas the analogous percentage for random traders is $26\%$.     

\begin{figure}
\begin{center}
\includegraphics[width=3.6in,angle=0]{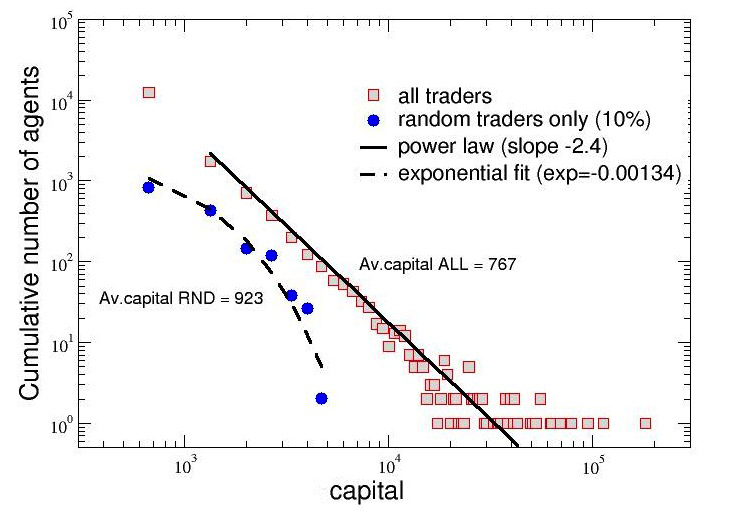}
\caption{\label{fig8} 
Comparison between the final capital/wealth distribution for the whole trading network with $10\%$ of random traders shown in Fig.7  (reported again as square symbols) and the same distribution for the random traders component only (full circles).
The latter can be fitted with an exponential distribution, also reported as dashed line, with an exponent equal to $-0.00134$. We also report the average capital calculated over all the traders (767 credits) and over
the random traders only (923 credits).See text for details.}
\end{center}
\end{figure}

These findings allow us to extend our previous results for single traders \cite{Biondo2} to collective effects in a community of  traders . In fact, they suggest  that the adoption of random strategies would diminish the probability of extreme events, in this case large  increases or  losses of wealth, but also ensure almost the same average wealth over a long time period, at variance with  technical strategies.
In particular, looking at the details of the two distribution shown in Fig.8, we also find that $40\%$ of RSI traders have a final capital smaller than the worst random trader, whereas only $3\%$ of RSI traders perform better than the best random trader. This means that, for technical traders, the risk of losses is much greater than the probability of gains, compared to those of random investors. Random trading seems therefore, after all, a very good combination of low risk and high performance.  

Before closing this section we note that all the numerical simulations presented are quite robust and that similar results were obtained adopting other historical time series, such as, for example, FTSE UK or FTSE Mib.

\section{Discussion and Potential Policy Implications}

Even though  our results were  obtained for a ``toy model'' of financial trading, we think that they have potentially interesting policy implications.

According to the conventional assumption, neither bubbles or crashes should occur when all agents are provided with the same complete and credible set of information about monetary and asset values traded in  the market. This  is the basis of the well-known Efficient Market Hypothesis \cite{Fama}, based on the paradigm of Rational Expectations \cite{muth}. Bubbles and crashes should also not occur according to the wide-spread Dynamic Stochastic General Equilbrium (DSGE) Models \cite{DSGE1,DSGE2,DSGE3,DSGE4}. However, the wisdom of crowds effect, which induces the equilibrium price, can be undermined by information feedbacks and social influence \cite{pnas-helbing-et-al}. Such social influence may lead to bubbles and crashes. 
\par
To account for herding effects, researchers have started to propose concepts such as ``rational bubbles'' \cite{Diba}, recognizing that it can be profitable to follow a trend. However, on average, trend-following is not more successful than random investments --- it is rather more risky \cite{Biondo2}. From human psychology, we know that people tend to follow others, i.e. to show herding behavior, if it is not clear what is the right thing to do \cite{cialdini}. This fact establishes that many traders will be susceptible to the trading decisions of others. It seems realistic to assume that each agent is endowed with a different quantity and quality of information, coming from private information sources with different reputation, depending on the agents' position in the network (see e.g. Refs. \cite{dellavignakaplan, cohen-frazzini-malloy, massa-simonov, goeree-palfrey-rogers-mckelvey, titman-trueman, beatty-ritter, michaely-shaw}).
However, such information feedbacks may be harmful, particularly when the market is flooded with volatile and self-referential information. 

Our paper supports the hypothesis \cite{Dirk2} that introducing ``noisy trading'' (i.e. random investors) in financial markets can destroy bubbles and crashes before they become large, and thereby avoid dangerous avalanches. By preventing extreme price variations, random investments also help to identify the equilibrium price \cite{teoh}.
It seems that already  a small number of random investors (relative to the total number of agents) would be enough to have a beneficial effect on the financial market, particularly if distributed at random. Such investors could be central banks, but also large investors, including pension funds or hedge funds with an interest in reducing the risks of their investments. 

We are aware that further studies with more sophisticated and realistic models of financial markets should be performed to explore the full potentials and limitations of random investment strategies. However, our results suggest that random investments will always reduce both the size and frequency of bubbles/crashes. Further research will be devoted to understanding  the most opportune timing for the introduction of such random investments and whether this innovative policy instrument can enable a smooth control of financial markets, thereby reducing their  fragility.

\section{Acknowledgments}

DH acknowledges partial support by the FET Flagship Pilot Project FuturICT (grant number 284709), the ETH project "Systemic Risks, Systemic Solutions" (CHIRP II project ETH 4812-1)  and the ERC Advanced Investigator Grant  "Momentum" (Grant No. 324247).
\vfill

\end{document}